\documentclass{moriond}

\usepackage{siunitx}

\def\be{\begin{equation}}
\def\ee{\end{equation}}
\def\bea{\begin{eqnarray}}
\def\eea{\end{eqnarray}}

\def\dd{\mathrm{d}}
\DeclareSIUnit \parsec {pc}

\newcommand{\Photo}{}

\begin{document}
\begin{flushright}
       MS-TP-23-17\\
\end{flushright}
\vspace*{4cm}
\title{Radiative corrections to stop-antistop annihilation}

\author{Luca Paolo Wiggering}

\address{Institut f\"ur Theoretische Physik, Westf\"alische Wilhelms-Universit\"at M\"unster, Wilhelm-Klemm-Stra{\ss}e 9, D-48149 M\"unster, Germany}

\maketitle\abstracts{
We compute the full $\mathcal{O}(\alpha_s)$ corrections to stop-antistop annihilation into two gluons and a light quark-antiquark pair within the framework of the Minimal Supersymmetric Standard Model (MSSM), including the non-perturbative Sommerfeld enhancement effect. Numerical results for the total annihilation cross section are shown and the effect on the neutralino relic density is discussed for an example scenario in the phenomenological MSSM. }

\section{Introduction}
\label{sec:intro}
Large-scale structure simulations of the Universe and astrophysical observations provide strong evidence for the existence of cold dark matter (CDM). Through the measurement of the temperature anisotropies in the Cosmic Microwave Background by the \emph{Planck} mission \cite{Planck:2018vyg}, the associated relic density of CDM can be constrained to the very narrow range
\be
    \Omega_{\rm CDM}h^2 ~=~ 0.120 \pm 0.001
    \label{eq:omh2}
\ee
at $1\sigma$ confidence level, where $h$ denotes the present Hubble expansion rate $H_0$ in units of $\SI{100}{\kilo\meter\per\second\per\mega\parsec}$. One possible particle physics explanation is that CDM consists of weakly interacting massive particles (WIMPs), as this 
scenario can not only explain the observed relic density, known as the WIMP miracle, but is also motivated by many extensions of the Standard Model (SM) addressing other open questions besides dark matter. As a byproduct, these models then predict the existence of new particles with masses around the electroweak scale which could be CDM. One of these extensions is the $R$-parity conserving MSSM, which provides with the lightest neutralino $\tilde{\chi}^0_1$ a stable WIMP.

As supersymmetry (SUSY) only adds new particles, but leaves the interaction strengths at the same order of magnitude, it is natural to assume that today's neutralino abundance is produced via thermal freeze-out, which is known to occur within the SM. The fact that due to $R$-symmetry every SUSY particle must eventually decay to the lightest neutralino allows then to write down the single Boltzmann equation 
\begin{equation}
    \dot{n}_\chi= -3 H n_\chi - \langle \sigma_{\mathrm{eff}} v\rangle \left(n^2_\chi-(n_\chi^{\mathrm{eq}})^2\right),
        \label{eq:boltzmann}
\end{equation}
for the neutralino number density $n_\chi$ with $n_\chi^{\mathrm{eq}}$ denoting the corresponding equilibrium value and $H$ the Hubble parameter \cite{Edsjo:1997bg}. Here, particle physics enters through the thermally averaged effective annihilation cross section 
\begin{equation}
    \langle \sigma_{\mathrm{eff}} v\rangle = \sum_{i,j} \langle\sigma_{ij} v\rangle \frac{n^{\mathrm{eq}}_i}{n_\chi^{\mathrm{eq}}} \frac{n^{\mathrm{eq}}_j}{n_\chi^{\mathrm{eq}}},
    \label{eq:eff_XSec}
\end{equation}
which is a sum over all possible annihilation channels of SUSY particles into SM final states. The Boltzmann suppression factor 
\begin{equation}
    \frac{n^{\mathrm{eq}}_i}{n_\chi^{\mathrm{eq}}} \sim \exp\left(-\frac{m_i-m_\chi}{T}\right)
\end{equation}
with the photon temperature $T$ gives the insight that besides pure neutralino annihilation, only those (co)annihilation channels with a small mass difference between at least one of the initial particles and the neutralino can contribute significantly to the relic density. 

As a necessary ingredient for the stabilisation of the Higgs mass against radiative corrections is a light stop, the lightest superpartner $\tilde{t}_1$ of the top quark is a theoretically well-motivated candidate for the next-to-lightest supersymmetric particle (NLSP). 
Another reason to consider light stops or another SUSY particle close in mass to the neutralino in general is that reconciling the increasing experimental neutralino mass limits with the observed dark matter density requires a mechanism to lower the relic density. This mechanism could e.g. be given by (co)annihilation.  

Even though the integration of the Boltzmann equation in Eq. \ref{eq:boltzmann} as well as the calculation of the associated (co)annihilation cross sections for generic new physics models is nowadays performed in a highly automatised fashion by public codes like MicrOMEGAs \cite{Belanger:2001fz} or MadDM \cite{Ambrogi:2018jqj}, these tools calculate the annihilation processes only at an (effective) tree level. This raises then the question whether theoretical predictions for how much dark matter there is in the universe can compete in precision with the experimental value from Eq. \ref{eq:omh2}. Building upon earlier works, e.g. \cite{Harz:2014tma,Schmiemann:2019czm}, we therefore study the impact of the full next-to-leading order (NLO) SUSY-QCD corrections to the processes $\tilde{t}_1 \tilde{t}^\ast_1 \to g g$ and $\tilde{t}_1 \tilde{t}^\ast_1 \to q \bar{q}$ on the neutralino relic density with $q$ denoting an effectively massless quark from the first or second generation. These two processes have been implemented simultaneously into the software package DM@NLO as they have to be combined at one-loop order in order to obtain an infrared safe cross section.

\section{Computational details of the SUSY-QCD corrections}
\begin{figure}
    \centering
        \centering
        \includegraphics[width=0.15\textwidth]{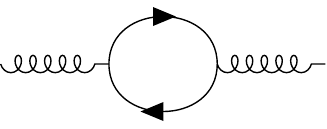} \ \ \ \
        \includegraphics[width=0.15\textwidth]{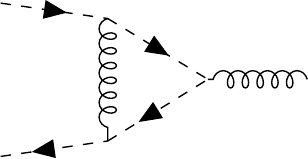} \ \ \ \
        \includegraphics[width=0.15\textwidth]{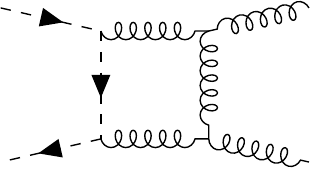} \ \ \ \
        \includegraphics[width=0.15\textwidth]{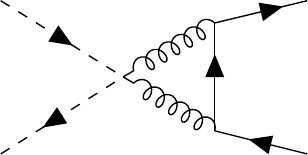} \ \ \ \
        \includegraphics[width=0.15\textwidth]{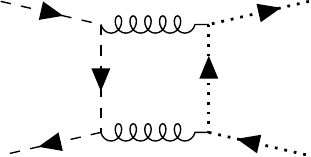} \ \ \ \
    \caption{A few example Feynman diagrams for the virtual corrections to stop annihilation.}
    \label{fig:diags_virtual}
\end{figure}

\begin{figure}
    \centering
        \includegraphics[width=0.14\textwidth]{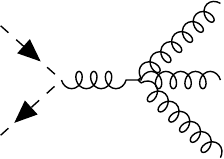} \ \ \
        \includegraphics[width=0.14\textwidth]{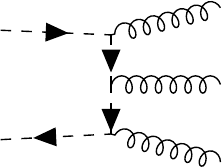} \ \ \
        \includegraphics[width=0.14\textwidth]{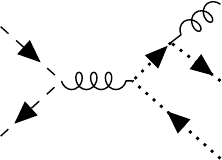} \ \ \ 
        \includegraphics[width=0.14\textwidth]{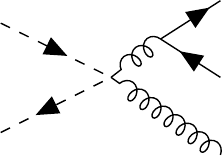} \ \ \
        \includegraphics[width=0.14\textwidth]{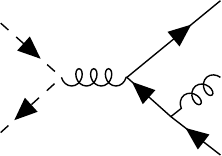} \ \ \
    \caption{A few example Feynman diagrams for the real corrections to stop annihilation.}
    \label{fig:diags_real}
\end{figure}

The NLO cross section 
\begin{equation}
    \sigma^{\mathrm{NLO}} = \sigma^{\mathrm{Tree}} + \Delta\sigma^{\mathrm{NLO}}
\end{equation}
in the strong coupling $\alpha_s$ consists of the tree-level part $\sigma^{\mathrm{Tree}}$ and the NLO correction 
\begin{equation}
   \Delta\sigma^{\mathrm{NLO}} = \int_2 \dd\sigma^{\mathrm{V}}+\int_3\dd\sigma^{\mathrm{R}} ,
\end{equation}
which is decomposed further into virtual $(\dd\sigma^{\mathrm{V}})$ and real corrections $(\dd\sigma^{\mathrm{R}})$ where the subscripts on the integrals refer to the number of final-state particles. As we use the metric tensor $g^{\mu\nu}$ to sum the two gluon polarizations, the unphysical longitudinal degrees of freedom of the external gluons are subtracted by including Faddeev-Popov ghosts as asymptotic states. In Figs. \ref{fig:diags_virtual} and \ref{fig:diags_real} a few example Feynman diagrams for both kinds of corrections are displayed where ghosts are shown as dotted lines with arrows indicating the ghost flow. 
The singularities occurring in the loops and within the three-particle phase space integration are regularised by working in $D=4-2\varepsilon$ dimensions following the SUSY-preserving four-dimensional helicity scheme  \cite{Signer:2008va}. The ultraviolet divergences are then cancelled through the renormalization of fields, masses and couplings. In order to make the phase space integration of the real emission matrix element numerically feasible, the Catani-Seymour dipole subtraction method for massive initial-states is employed \cite{Catani:1996jh,Harz:2022ipe}. This subtraction method is based on the construction of an auxiliary cross section $\dd\sigma^{\mathrm{A}}$ that cancels the infrared singularities in the real-emission matrix element pointwise and is at the same time simple enough so that the integration over the singular region can be performed analytically, allowing in turn the cancellation of the infrared poles contained in the virtual part. This procedure can be summarised symbolically as  
\begin{equation}
   \Delta\sigma^{\mathrm{NLO}}=\int_{3}\left[\dd{\sigma}^{\mathrm{R}}_{\varepsilon=0}-\dd\sigma^{\mathrm{A}}_{\varepsilon=0}\right]\\ +\int_{2}\left[\dd{\sigma}^{\mathrm{V}}+\int_{1}\dd\sigma^{\mathrm{A}}\right]_{\varepsilon=0}.
\end{equation}

Besides the one-loop corrections in the strong coupling, another piece in the calculation comes from the exchange of potential gluons between the incoming stop-antistop pair with each gluon contributing a factor $\alpha_s/v$. For small relative velocities $v$ typical for the freeze-out regime, these contributions become non-perturbative and have to be resummed up to all orders. This is known as the Sommerfeld enhancement effect \cite{Sommerfeld:1931qaf} and an analytical treatment is achieved via the framework of non-relativistic QCD \cite{Kiyo:2008bv}, giving the improved cross section $\sigma^{\mathrm{Som}}$. The full cross section $\sigma^{\mathrm{Full}}$ is obtained by matching the fixed-order calculation to the resummation calculation. It is also possible to subtract the velocity enhanced part from the virtual corrections giving the "pure" NLO correction $\sigma^{NLO}_v$. More computational details are available in the original publication \cite{Klasen:2022ptb}. For the investigation of bound state formation in a simplified dark matter model similar to the stop and neutralino sector of the MSSM, we refer to \cite{Becker:2022iso}. 

\section{Impact on the annihilation cross section and the relic density }
\begin{figure}[h]
\centering
\includegraphics[width=0.48\textwidth]{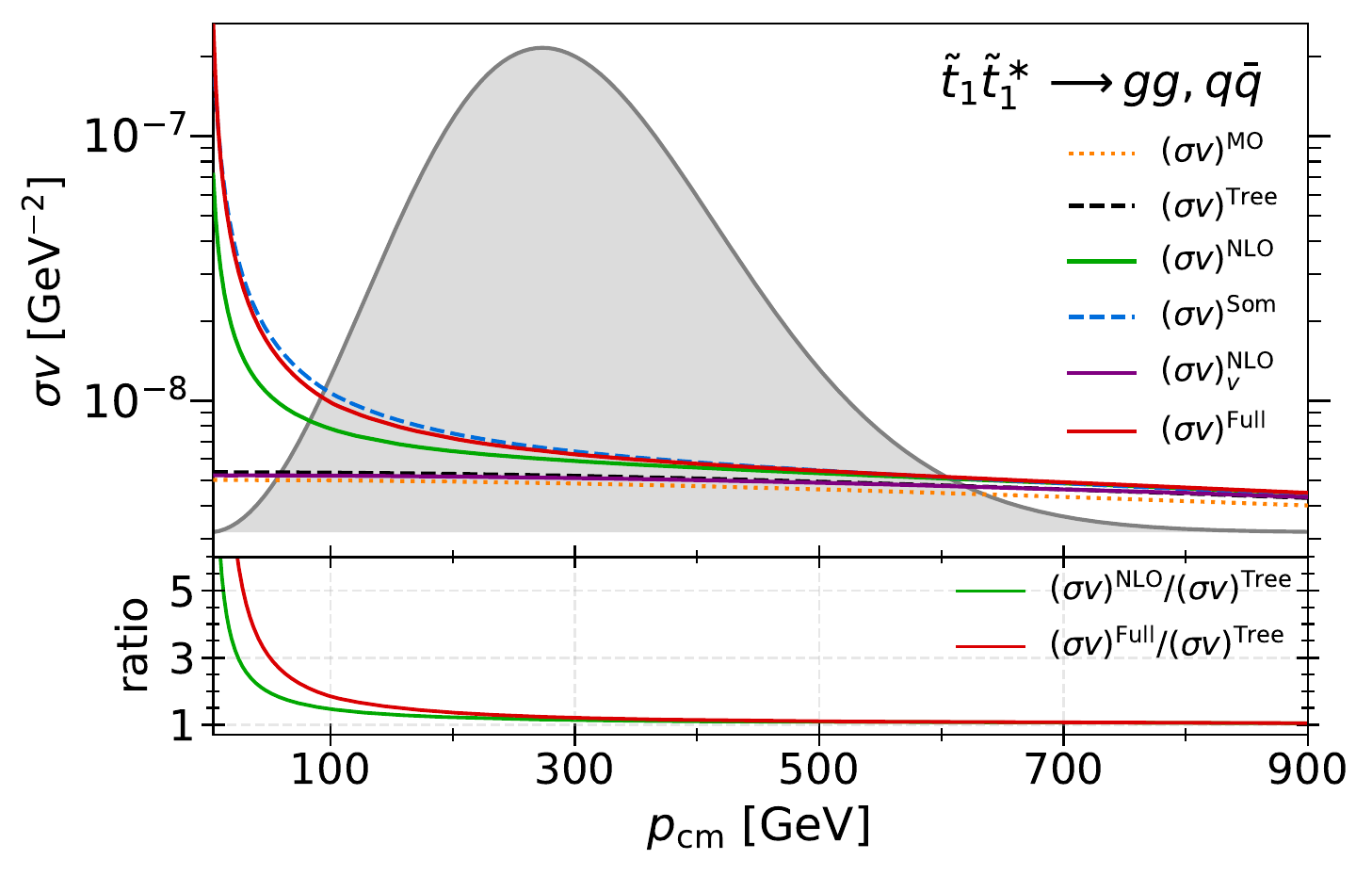}
\includegraphics[width=0.48\textwidth]{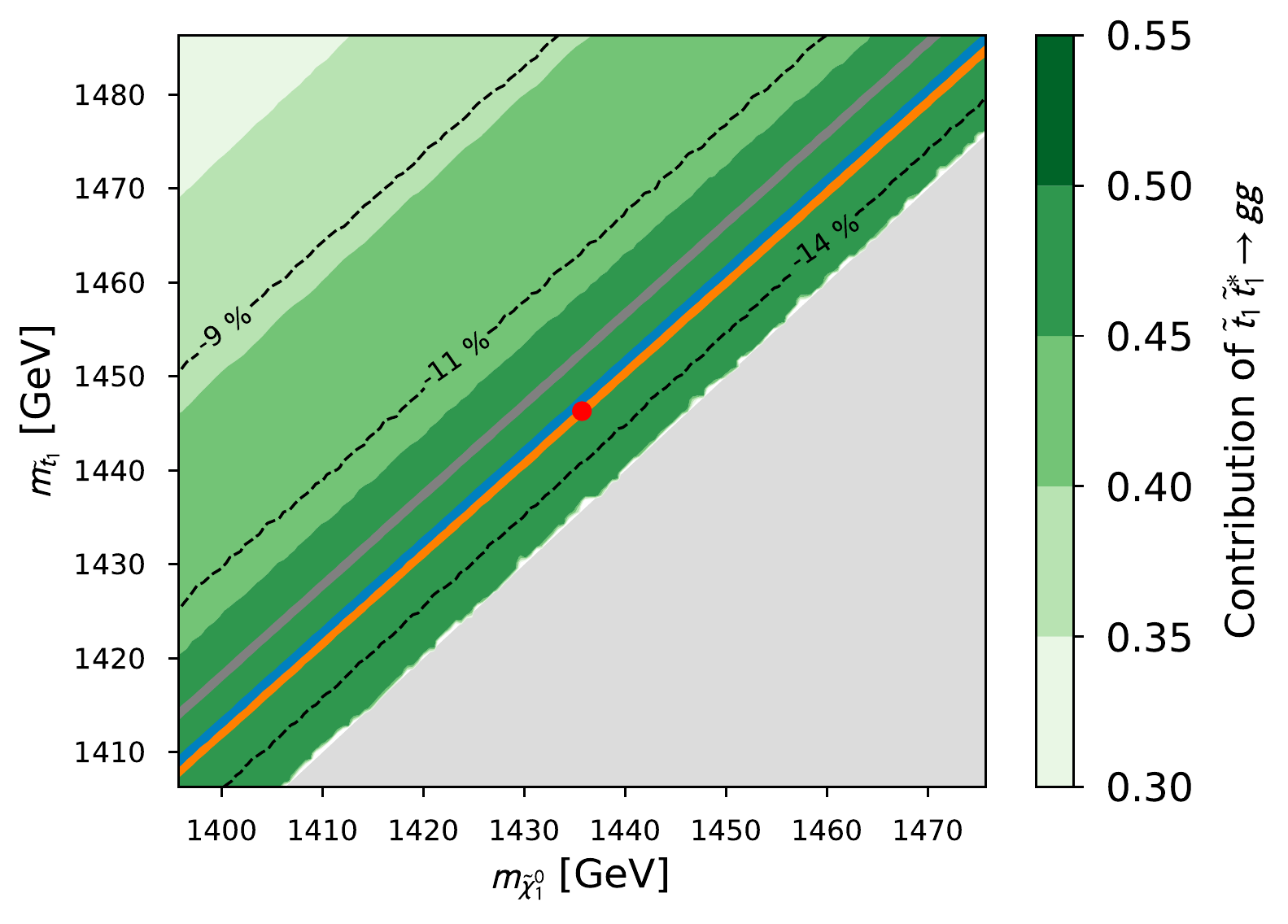}
\caption{Left: Annihilation cross section times velocity $\sigma v$ for stop-antistop annihilation into gluons and light quarks. The leading order cross section obtained with CalcHEP is denoted by $\sigma^{\mathrm{MO}}$. The grey region corresponds to the Maxwell-Boltzmann distribution entering the thermal average at the freeze-out temperature in arbitrary units. Right: Contribution of stop annihilation into gluons to the relic density in the neutralino and stop mass plane. The parameter region consistent with the observed relic density at the $2\sigma$ confidence level based on the MicrOMEGAs 2.4.1 calculation is shown in orange, in blue for the DM@NLO
tree-level result and in grey for the fully corrected cross section. The red dot represents the chosen scenario and the black lines indicate the shift in the relic density compared to the tree-level result.}
\label{fig:impact}
\end{figure}

In Fig. \ref{fig:impact} the impact of the radiative corrections on the annihilation cross section as well as on the cosmologically preferred parameter region in the physical neutralino and stop mass plane is
shown for a currently viable pMSSM-19 scenario. In this scenario the mass difference between the bino-like neutralino and the stop is $m_{\tilde{t}_1}-m_{\tilde{\chi}^0_1}=\SI{10.6}{\giga\electronvolt}$, so that stop annihilation into gluons is with $\SI{47}{\percent}$ the dominant contribution to the relic density followed by stop pair-annihilation into top quarks with $\SI{23}{\percent}$. The full cross section is in very good approximation given by the Sommerfeld enhancement alone since the pure NLO corrections without the velocity enhanced part are below $\SI{\pm 3}{\percent}$. However, the full corrections are still large enough to shift the relic density beyond the uncertainty of the Planck measurement. In order to compensate the increased effective annihilation cross section, the stop mass needs to be lifted by $\SI{6.1}{\giga\electronvolt}$ compared to the MicrOMEGAs calculation. For a more detailed discussion of the chosen scenario and the numerical results also for the additional inclusion of the full one-loop corrections to stop pair-annihilation and neutralino-stop co-annihilation we refer to \cite{Klasen:2022ptb}.

\section{Conclusion}
We presented the impact of the full one-loop SUSY-QCD corrections on the stop annihilation cross section into gluons and light quarks, including the Sommerfeld enhancement effect. We find that these corrections are crucial as they can shift the value of the relic density beyond the current experimental uncertainty for the example SUSY scenario. However, the major correction can be captured through the Sommerfeld enhancement alone. We are also confident that this approximation extends to general dark matter models containing coloured scalars. 

\section*{Acknowledgments}
This work has been supported by the Deutsche Forschungsgemeinschaft (DFG, German Research Foundation) through the Research Training Group GRK 2149. L.W. thanks the organisers of Moriond EW 2023 for the great conference.

\end{document}